\documentclass[letterpaper,twocolumn,10pt]{article}
\usepackage{usenix2019,epsfig,endnotes}
\usepackage[dvipsnames]{xcolor}

\usepackage{hyperref}
\usepackage{graphicx}
\usepackage{comment}
\usepackage{authblk}
\graphicspath{ {./images/} }

\title{\LARGE \bf
Anti-Turing Machine
}

\author[1]{Viacheslav Dubeyko}

\begin{document}

\maketitle
\pagestyle{plain}

%%%%%%%%%%%%%%%%%%%%%%%%%%%%%%%%%%%%%%%%%%%%%%%%%%%%%%%%%%%%%%%%%%%%%%%%%%%%%%%%
\begin{abstract}

The invention of CPU-centric computing paradigm was incredible breakthrough of computer science that revolutionized our everyday life dramatically. However, the CPU-centric paradigm is based on the Turing machine concept and, as a result, expensive and power-hungry data transferring between the memory and CPU core is inevitable operation. Anti-Turing machine paradigm can be based on two fundamental principles: (1) data-centric computing, and (2) decentralized computing. Anti-Turing machine is able to execute a special type of programs. The commands of such program have to be addressed to the 2D or 3D persistent memory space is able to process data in-place. This program should not define the position or structure of data but it has to define the goal of data processing activity. Generally speaking, it needs to consider the whole memory space like the data transformation space.  But the data placement, particular algorithm implementation, and strategy of algorithm execution are out of scope of the program.

\end{abstract}

%%%%%%%%%%%%%%%%%%%%%%%%%%%%%%%%%%%%%%%%%%%%%%%%%%%%%%%%%%%%%%%%%%%%%%%%%%%%%%%%


%%%%%%%%%%%%%%%%%%%%%%%%%%%%%%%%%%%%%%%%%%%%%%%%%%%%%%%%%%%%%%%%%%%%%%%%%%%%%%%%
\section{INTRODUCTION}

\textbf{Turing machine} represents the hypothetical machine that was invented by Alan Turing in 1936 year. Theoretically, this machine is able to simulate a computer algorithm of any complexity. The Turing machine includes an “infinite” tape that works like computer memory or data storage. Such “infinite” tape is split on positions and every position can keep one symbol. The next important item of Turing machine is a read-write head that points out a particular position of “infinite” tape at every particular time point. As a result, Turing machine is able to read the symbol in the current position and this symbol works like the code of operation. The symbol defines the behavior of Turing machine. Finally, such sequence of steps can be executed till the algorithm’s end, the tape’s end or encountering error (or unknown symbol) in the algorithm.

\textbf{The nature of data} is complex with multiple dimensions, relations, and dependencies. Moreover, data is always continuously evolving. However, Turing machine is the pure example of algorithm-oriented approach. Generally speaking, the algorithm-oriented paradigm implies by definition the presence of one processing core that executes the algorithm sequentially. The fundamental issue of CPU-centric paradigm is the inevitable transferring as data as code to the place of processing. The processing core is able to process only one data portion for the single clock tick but not the whole data array at once.

\textbf{Traditional model of data processing} distinguishes data and request. These entities (data and request) are fundamentally different information streams. It is possible to say that data contains the internal structure that needs to be recognized. Moreover, any data represents a request that a system needs to recognize. Such immanent request is the basis for elaboration of system reaction by means of synthesis of knowledge or program code.

\textbf{The fundamental feature of data-centric computing paradigm} is the possibility to distribute data processing in the whole persistent memory space without the necessity to deliver data to the centralized processing core for the data transformation. Data-oriented paradigm means that data is merged with processing units. Generally speaking, the processing space can be represented like 2D or 3D array of processing units where each Processing Unit (PU) has dedicated capacity of the persistent memory (able to store some piece of data). Anti-Turing machine considers every Processing Unit (PU) like the independent item that makes own decision about starting and ending point of execution independently. In other words, PU participates in data processing only if PU's data is satisfied to the conditions of a requested operation. Data Processing Unit (DPU) is the active entity is capable to process the data and to interact, collaborate with other DPUs. The goal of such model is to implement the decentralized model of data processing. The decentralized model implies that nature of data defines where and what algorithm will be applied in the whole persistent space instead of centralized algorithm that needs to be executed by centralized core. The initiator injects the request into the persistent space but it doesn't define the place and algorithm of data processing. Generally speaking, the persistent space has the active nature by virtue of the capability of every DPU to make the decision to execute or to ignore an initiator's request independently. As a result, DPU array defines internally the distribution of activity by data processing in the case of receiving a request.

\section{The Nature of Conflict between Algorithm-oriented and Data-oriented Computing}

\textbf{Algorithm-oriented computing}. The nature of data is complex with multiple dimensions, relations, and dependencies. Moreover, data is always continuously evolving. But any algorithm expects an one-dimensional and simple structure that is stopped to evolve. It is possible to say that any algorithm would like to see the exclusive access to the data without any modifications by anyone else. Even if it is possible to use the different synchronization primitives (semaphores, mutexes, and so on) for implementation protocols of consistent modification and access of shared data but, anyway, the algorithm-oriented paradigm creates a lot of issues for easy achieving the inconsistent state of shared data in the multi-threaded environment. Usually, the native approach of algorithm-oriented paradigm is to lock the whole complex structure and to process the structure the step by step in iterative manner. Generally speaking, the algorithm-oriented paradigm implies by definition the presence of one processing core that executes the algorithm sequentially. The responsibility of the system is to transfer the data/code from the persistent storage or the input device near to the core for processing/execution. Even if the different time slices of the same algorithm can be executed by physically different cores but an algorithm sees the same virtual core. And the whole execution paradigm of data processing is rotating around this virtual execution core. It means that data can be processed only inside the execution core and the data transferring process is inevitable step. Turing machine has only one read/write head with the finite controller that results in the necessity to move some tape’s position for the processing. Generally speaking, the fundamental issue of CPU-centric paradigm is the inevitable transferring as data as code to the place of processing. The processing core is able to process only one data portion for the single clock tick but not the whole data array at once.

\noindent
\textbf{Data-oriented computing}. The goal of data-centric paradigm is opposite to the algorithm-oriented approach. The data-centric approach has to provide the opportunity to process the whole data array at once in the environment of multiple relations among data and continuous data evolution. From one point of view, an algorithm always implies some data structure because any algorithm is dependent from the data structure. However, the nature of algorithm limits any developer by abstraction of data structure. It is possible to say that the most natural data structures are array and list for any algorithm. Finally, the one core (or read/write head) needs to work with the whole data structure. Data-centric paradigm implies the completely opposite opportunity, namely, to distribute the transformation cores in the persistent data space. Generally speaking, the extreme case could be represented like every piece of data has own dedicated processing core. Such data-centric approach is completely, fundamentally incompatible with algorithm-oriented paradigm. First of all, the delivery of code to the every processing core could be very expensive for the case of data-centric paradigm if the Turing machine remains the cornerstone execution approach. Secondly, the data-centric paradigm provides the opportunity to modify the every data item independently instead of necessity to manage the whole data structure (for the case of algorithm-oriented approach). Finally, data-centric computing paradigm is able to provide the flexible way of continuous processing of evolving data in the real-time manner. Generally speaking, data-centric computing implies that particular elementary core is responsible for independent definition/selection of processing algorithm is applied on dedicated data portion. However, the host needs to elaborate a strategy that orchestrates the data processing activity in the matrix of processing cores.

\noindent
\textbf{CPU-centric vs. Data-oriented computing}. Algorithm-oriented paradigm sounds like management and data flows need to achieve one centralized computing point where the real transformation can take place. Oppositely, data-centric computing paradigm can be represented like the distributed array, matrix or space of computing points are able to process data. Generally speaking, algorithm-oriented computing implies the scheduling is distributed in time. It means that scheduler needs to distribute the CPU's time slices in the case of algorithm-oriented computing. The data-centric computing creates completely different scheduling paradigm. It creates the opportunity to distribute the data processing in 2D or 3D computing space. Finally, it sounds that scheduler needs to distribute the computing not in time but in space. Every computing core in data-centric paradigm is independent from another ones and it is able to elaborate an own algorithm of data processing on the basis of data type (or other knowledge) without the necessity to deliver the code to the place of data processing. The key peculiarity of data-centric computing is that independent, active processing cores own by piece of data are able to accept and to manage the independent evolution and modification of different piece of data. Generally speaking, the active processing cores are capable to react on data portions’ independent evolution by means of rebuilding the relations or reworking the knowledge about existing data. The 2D/3D matrix of active and independent processing cores creates the really important point because these distributed active cores are able to interact, to create relations, to build a new knowledge, to collaborate, or to compete with each other. Finally, the data-centric computing paradigm is capable to create the “alive” computing system that will evolve together with data by means of building a new knowledge and elaboration the behavioral strategy.

\section{The Importance of Data-oriented Computing for Next Generation NVM Memory}

It is possible to say that the block-based interface was the single way of accessing the persistent data for some time. The reason of using the block-based interface took place because of very slow nature of persistent storage technologies. Also the peculiarities of technologies of persistent memory (for example, PMR or NAND flash) dictate the necessity to use a physical sector’s size like the base granularity of data exchange between the host and persistent storage device. Generally speaking, the necessity to transfer the data from the storage space to the CPU core (processing place) was inevitable side effect. However, the invention of CPU-centric computing paradigm was incredible breakthrough of computer science that revolutionized our everyday life dramatically. But nowadays the volume of existing data is huge and growing exponentially. The reality of Big Data suffers from the lack of necessary computing power that takes place because of widely used CPU-centric computing paradigm. Moreover, the next generation of NVM memory is byte-addressable, persistent memory that, theoretically, is capable to increase the available computing power dramatically. However, the NVM memory has the fundamental contradiction with the CPU-centric computing paradigm. Generally speaking, theoretically, a byte-addressable and persistent memory is able to decrease the distance between processing core and data placement till zero distance. But CPU-centric paradigm is based on the Turing machine concept and, as a result, expensive and power-hungry data transferring between the memory and CPU core is inevitable operation. It is possible to conclude that CPU-centric computing paradigm is exhausted and obsolete concept that nowadays is the crucial bottleneck in the direction of increasing the computing power. Another very critical problem is the enormous volume of existing data that is continuously growing. The needs in processing of Big Data is very challenging, time-consuming, and power-hungry problem. As a result, the power consumption on data transferring during the Big Data processing is unaffordable luxury. Moreover, algorithm-oriented nature of modern computational model becomes the critical drawback that prevents the data processing to be deeply distributed. The key reason of this issue is the primacy of algorithm in the modern computing paradigm. Generally speaking, data is treated like passive media that needs to be processed by an algorithm. However, data has own nature that can be "elaborated" and "realized" by a computing core. It means that discovered nature of data can be the basis for data processing without the necessity to use the algorithms are written by people. Byte-addressable NVM memory is the first step in the direction of invention of a new computing paradigm. The responsibility of the new computing paradigm is to build a computing environment that provides opportunity to synthesize a processing algorithm by data itself with the goal to achieve the maximum possible distributed and decentralized data processing in the persistent space of memory itself. Recent advantages in AI and neuromorphic computing areas make real the goal to synthesize the algorithm on the basis of data nature. It means that if an algorithm can be synthesized and can be applied in the persistent memory space then the step of data transportation between the memory and CPU core can be completely eliminated. Moreover, data processing will be deeply decentralized and distributed.

\section{The Importance of Data-oriented Computing for AI Problem}

AI problem needs in proper building blocks. If anybody will consider the neuromorphic computing like a potential solution of the AI problem then it is clear that the computing paradigm cannot be algorithm-oriented. Only data can be the fundamental basis of neuromorphic computing because this paradigm has to synthesize or to elaborate an algorithm of data processing. Another valuable point could be the functional model of human (or animal) brain. It is possible to state that basic working units of brain (neurons) keep some portions of data and they join into a network with the goal to cooperatively process the data. Generally speaking, brain’s functional model is data-oriented but not the algorithm-oriented. As a result, AI problem needs in data-oriented computing paradigm. Usually, computing system represents the data as binary stream that is contained by file of "infinite" length. Nowadays, file concept is the unaffordable luxury that doesn't provide a good ecosystem in the Big Data world. The file hides, eliminates the data’s structure. It makes the data by "white noise" that completely destroys the knowledge about existed data nature. The conversion of data into the digital "white noise" becomes the very critical issue for AI problem. Generally speaking, the knowledge of data nature has to exist in the persistent space of data storage. The necessity to transfer the data to the place where the application has the knowledge about data nature (data structure) makes the whole system like the one big bottleneck. This bottleneck is trying to return our world in the Stone Age. The AI architecture needs to be based on the fundamental concept that elementary "atoms" of data is capable to keep the knowledge about own nature and is able to integrate into structures in the persistent memory space. The data-oriented paradigm's implementation can be imagined like multiple elementary and simple Processing Units (PU) are placed (or integrated) directly into the persistent NVM memory space. The elementary PU can be implemented by different ways. But one of the smallest possible implementation can be merging the memory cell with PU's functionality. Such union of memory cell and PU's functionality (arithmetical and logical operations) can be built on the physical phenomena of a media or special circuitry architecture of the memory cell. As a result, the combination of memory cell with arithmetical and logical operations functionality can be imagined like the elementary neuron is able to store and to process the elementary data portion. Generally speaking, the elementary neurons are able to combine into various structure if the every neuron has the knowledge about nature of stored data. It is possible to point out that the process of data registration is the very important step because the system is able to create the knowledge of data nature during the registration process.

\section{Dualistic Nature of Data}

Traditional model of data processing distinguishes data and request. These entities (data and request) are fundamentally different information streams. Usually, data is a binary stream that needs to be processed by some algorithm(s). It means that data is always the passive substance, from the current computing paradigm point of view. Oppositely, the request is the active substance that can be represented by some algorithm or sequence of instructions. The request can be recognized and be executed by CPU, for example. Generally speaking, the fundamental point is that the request represents a sequence of keywords defining the essence of algorithm. Any deviation or anomaly in the sequence of instructions results in the incorrect algorithm's behavior and introduces the bug. If anyone considers any data outside the scope of algorithm then it is possible to say that the data is binary stream or "white noise". An algorithm is able to recognize the structure of the data and to process the data in some way. However, any data in our life is not the "white noise". Information is the data only if it is possible to recognize the structure or information organization. Namely, recognizable logical structure is the immanent characteristic of data. Generally speaking, transformation of data into the binary (digital) form loses the very important and immanent features of data. It means that current computing paradigm drops the logical structure of data by means of storing data in the binary stream. Moreover, any data contains not only information but also the request's items. From one point of view, data is able to represent some abstract formula of algorithm or description that can be converted into the algorithm's implementation. But, from another point of view, any data is stored into the system or data is interacting with the system defines the relevant keywords or actual directions that creates the scope of knowledge extraction and evolving the knowledge in the system. It is very important to point out that data has immanent internal structure. Usually, this internal structure is defined by a registration method. Finally, it means that the immanent data's structure can be the steady basis for self-organization of data. Generally speaking, the using of internal nature of data could be a basis for data organization without the necessity to use the file system paradigm, for example. The self-organization of data is the powerful mechanism for the knowledge synthesis, the synthesis of program code, or self-evolving of the system. It is possible to say that data contains the internal structure that needs to be recognized. Moreover, any data represents a request that a system needs to recognize. Such immanent request is the basis for elaboration of system reaction by means of synthesis of knowledge or program code. Finally, the dualistic nature of data is very important point for the case of decentralized data processing model.

\section{Anti-Turing Machine Paradigm like the Basis of Data-oriented Computing}

\textbf{The key bottleneck of Turing machine paradigm} is fundamental algorithm-oriented nature. Nowadays, the evolution of computing technologies encounters the necessity to process much bigger volumes of data that it was a decade ago. Moreover, the data could be located as locally as on remote systems. As a result, the algorithm-oriented nature of Turing machine dictates the inevitable necessity to transfer data from the memory device (DRAM, storage device) to the place of data processing and backwards. It is possible to conclude that growing volume and deeply distributed nature of data requires a new computing paradigm. One of the possible direction of computing paradigm changing could be moving from algorithm-oriented to the data-centric computing paradigm.

\noindent
\textbf{The fundamental feature of data-centric computing paradigm} is the possibility to distribute data processing in the whole persistent memory space without the necessity to deliver data to the centralized processing core for the data transformation. Anti-Turing machine could be an implementation of data-oriented computing. The key reason of data transfer between the memory and processing core for Turing machine concept is by virtue of keeping knowledge about data structures in the application's processing algorithm. Generally speaking, the algorithm keeps the knowledge about structure of data that is represented like binary stream in the persistent memory. It means that Anti-Turing machine paradigm needs to store the knowledge about data structure in the persistent memory space. The data should be represented not by binary stream but it needs to store the knowledge about nature and structure with the data. The key goal of this concept is the opportunity to offload the data processing in the persistent memory, to distribute the data transformation through the whole storage space, and to prepare the infrastructure for embedding the AI primitives into the persistent memory.

\noindent
\textbf{Abstract machine concept}. If anybody considers the Turing machine paradigm then the fundamental basis of this paradigm is the sequence of instructions and every instruction define the data and the operation for this step of an algorithm. It means that the instruction code defines: (1) placement of data item (or array of items), (2) granularity of data item, and (3) machine code is capable be executed by CPU's core. The Anti-Turing machine paradigm needs to exclude the necessity to define the data placement, the data granularity, and the code of micro-program of the CPU's core. The new paradigm needs to get rid of the paradigm of centralized execution of algorithm (the fundamental basis of Turing machine). Now we are using the various high-level programming languages. These languages hide many details of CPU's internals and to provide the opportunity to develop the program for an abstract machine. It is possible to imagine that Anti-Turing machine is able to be based on the same principle. It means that Anti-Turing machine is able to execute a special type of programs. The commands of such program have to be addressed to the 2D or 3D persistent memory space is able to process data in-place. This program should not define the position or structure of data but it has to define the goal of data processing activity. Generally speaking, it needs to consider the whole memory space like the data transformation space.  But the data placement, particular algorithm implementation, and strategy of algorithm execution are out of scope of the program. In the new computing paradigm, every program code is a set of keywords that define: (1) some set of data for applying an algorithm, (2) condition(s) of applying the algorithm, (3) abstract or generalized algorithm definition.

\noindent
\textbf{Data-oriented paradigm} means that data is merged with processing units. Generally speaking, the processing space can be represented like 2D or 3D array of processing units where each PU has dedicated capacity of the persistent memory (able to store some piece of data). Anti-Turing machine considers every PU like the independent item that makes own decision about starting and ending point of execution independently. In other words, PU participates in data processing only if PU's data is satisfied to the conditions of a requested operation. First of all, every PU has to know the features of stored data (for example, keywords). The known features are the basis for involvement a particular PU into data processing activity. Generally speaking, it is the basis for transformation of abstract request into the real algorithm of data processing in the particular processing core. The requested features (for example, keywords) in an abstract request define what processing core will react by means of transformation of stored data. Also these features are the basis for synthesis of an algorithm of data processing in the particular processing core. Such generalization provides the opportunity to evolve by particular PU independently by means of synthesis of specialized algorithm for particular data. In other words, every PU implements own version of an algorithm for the stored data. This paradigm provides the flexibility as for data distribution in the persistent memory space as for synthesis of algorithm by different PU in the scope of the same abstract request. Generally speaking, the request initiator doesn't define the PU and data that have to be processed by the request.  A generalized request is addressed to the whole persistent processing space and this request can be delivered to every PU like broadcast message or to be distributed by means of special routing policy. The isolation of the initiator from the definition of execution place provides the flexibility as for easy evolution of persistent processing space as for synthesis of requests of any complexity without the necessity to modify the execution space.

\noindent
\textbf{DPU as finite automata}. The every item of persistent execution space (DPU) represents a finite automata that waits a request. If the DPU receives the request then it needs to detect that request contains any keyword is relevant to the data is stored in dedicated persistent memory. As a result, DPU makes the decision to ignore the request if there is no relevant keyword in request or, oppositely, to participate in request execution for the data in dedicated portion of persistent memory. One of the possible case could be the situation when all DPUs in the array have made the decision not to participate in the request processing by virtue of the absence of relevant data. How is request initiator able to know that the whole DPU array has rejected the request processing? Generally speaking, the Network-On-Chip (NOC) has to be responsible for managing this case. The network has responsibility to deliver the request from initiator to every DPU in the array following by some routing policy. It means that a packet with request should be transferred through the network achieving every DPU. As a result, DPU needs to analyze the request's packet, makes the decision to process the request or not, and, finally, to change the packet with the goal to inform the requester about the decision. Generally speaking, the packet could contain two special counters. One counter has to account the number of DPUs are processing the request but another one has to account the number of rejections. But if a DPU makes decision to process the request then it could send to the initiator a confirmation request with the DPU's identification and the forecast how soon the request could be finished. Finally, the initiator will be able to receive one or several confirmation packets or only the initial packet with the information about number of DPUs are working with the request. If the DPU array doesn't contain the relevant information then the initial packet will contain only number of rejections to process the request. The first step of any DPU has to be the identification of data that needs to be processed by a request. If some data was identified by the DPU then it needs to check that data is satisfied to condition of the request. As a result, DPU will prepare the data set (view) is suitable for the request. Generally speaking, this step could the end of DPU's activity if the initiator requested only data search. However, the next step could be the applying of some transformation algorithm on the prepared view. An algorithm can be synthesized or be selected from the existing set of functional blocks. Moreover, this algorithm can be identified by some keyword in the initial request.

\section{Decentralized Nature of Anti-Turing Machine Paradigm}

\textbf{The fundamental basis of Anti-Turing machine} is the model of DPU. Every DPU represents the union of PU and NVM memory. The responsibility and goal of NVM memory is to keep some portion of data and the knowledge about these data (for example, characteristic keywords of the data portion). Oppositely, the responsibility of PU is to interact with the "outer world" by means of providing some services for accessing and transformation the data on the basis of DPU's knowledge about data. The interaction of DPU with the "outer world" is able to increase the knowledge about data and to provide the opportunity to evolve of DPU's functionality. Generally speaking, DPU is the active entity is capable to process the data and to interact, collaborate with other DPUs. The goal of such model is to implement the decentralized model of data processing.

\noindent
\textbf{The decentralized model} implies that nature of data defines where and what algorithm will be applied in the whole persistent space instead of centralized algorithm that needs to be executed by centralized core. The initiator injects the request into the persistent space but it doesn't define the place and algorithm of data processing. Generally speaking, the persistent space has the active nature by virtue of the capability of every DPU to make the decision to execute or to ignore an initiator's request independently. As a result, DPU array defines internally the distribution of activity by data processing in the case of receiving a request. Moreover, the nature of stored data defines the distribution of data processing load in the DPU array. Finally, the active nature of DPU array is the key approach for implementation of decentralized model of data processing in Anti-Turing machine. The decentralized model of data processing is the key ingredient of Anti-Turing machine by virtue of the opportunity to isolate the particular implementation of DPU array (and evolution of DPU array) from the representation of initiator's requests. Generally speaking, decentralized model creates the way to evolve independently as for initiator's requests (abstractness and complexity) as for DPU array (internal organization and relations). It means that DPU array is able to evolve continuously in the direction of increasing the data volume, gathering the knowledge, and creation more and more complicated relations amongst the DPUs.

\noindent
\textbf{The evolution of DPU array} can take place without the necessity to inform the initiator or to affect the initiator's requests. Moreover, decentralized nature of data processing in DPU array doesn't need in centralized metadata structures because of every DPU plays the active and independent role in the data processing. It makes sense to point out that adding or deletion of data in the DPU array (like persistent storage) is the simple procedure that doesn't require the immediate creation or deletion of relations or links in the DPU array. Generally speaking, the creation or destruction of relations amongst the DPU is capable to take place gradually by means of slow evolution  without the necessity to use some centralized metadata storage. Moreover, growing knowledge about stored data could discover the new relations that wouldn't be created at the moment of adding data in the DPU array. The really important point of the decentralized model is the capability to create the relations internally in the DPU array without the participation of any external agent. It means that the evolution of DPU array is able to take place independently and with internal elaboration of relations between DPU. Generally speaking, the evolving nature of relations detection is capable to find the unexpected relations that cannot be predicted or be detected by strictly defined structure or initiator's request. Moreover, such evolution of DPU array creates the flexibility in relations creation and knowledge extraction on the basis of existing data in the DPU array. It is possible to state that internal relations creation in the DPU array is the basis for efficient and flexible policy of data processing distribution in the DPU array. Moreover, the evolving nature of DPU array is steady basis for code and circuitry synthesis in the DPU array that can provide the infrastructure for various AI approaches.

\section{Conclusion}

The invention of CPU-centric computing paradigm was incredible breakthrough of computer science that revolutionized our everyday life dramatically. But nowadays the volume of existing data is huge and growing exponentially. The reality of Big Data suffers from the lack of necessary computing power that takes place because of widely used CPU-centric computing paradigm. Moreover, the next generation of NVM memory is byte-addressable, persistent memory that, theoretically, is capable to increase the available computing power dramatically. However, the NVM memory has the fundamental contradiction with the CPU-centric computing paradigm. Generally speaking, theoretically, a byte-addressable and persistent memory is able to decrease the distance between processing core and data placement till zero distance. But CPU-centric paradigm is based on the Turing machine concept and, as a result, expensive and power-hungry data transferring between the memory and CPU core is inevitable operation. It is possible to conclude that CPU-centric computing paradigm is exhausted and obsolete concept that nowadays is the crucial bottleneck in the direction of increasing the computing power.

If anybody considers the Turing machine paradigm then the fundamental basis of this paradigm is the sequence of instructions and every instruction define the data and the operation for this step of an algorithm. It means that the instruction code defines: (1) placement of data item (or array of items), (2) granularity of data item, and (3) machine code is capable be executed by CPU's core. The Anti-Turing machine paradigm needs to exclude the necessity to define the data placement, the data granularity, and the code of micro-program of the CPU's core. The new paradigm needs to get rid of the paradigm of centralized execution of algorithm (the fundamental basis of Turing machine). Now we are using the various high-level programming languages. These languages hide many details of CPU's internals and to provide the opportunity to develop the program for an abstract machine. It is possible to imagine that Anti-Turing machine is able to be based on the same principle. It means that Anti-Turing machine is able to execute a special type of programs. The commands of such program have to be addressed to the 2D or 3D persistent memory space is able to process data in-place. This program should not define the position or structure of data but it has to define the goal of data processing activity. Generally speaking, it needs to consider the whole memory space like the data transformation space.  But the data placement, particular algorithm implementation, and strategy of algorithm execution are out of scope of the program.


%%%%%%%%%%%%%%%%%%%%%%%%%%%%%%%%%%%%%%%%%%%%%%%%%%%%%%%%%%%%%%%%%%%%%%%%%%%%%%%%

%%%%%%%%%%%%%%%%%%%%%%%%%%%%%%%%%%%%%%%%%%%%%%%%%%%%%%%%%%%%%%%%%%%%%%%%%%%%%%%%

%%%%%%%%%%%%%%%%%%%%%%%%%%%%%%%%%%%%%%%%%%%%%%%%%%%%%%%%%%%%%%%%%%%%%%%%%%%%%%%%

%%%%%%%%%%%%%%%%%%%%%%%%%%%%%%%%%%%%%%%%%%%%%%%%%%%%%%%%%%%%%%%%%%%%%%%%%%%%%%%%

\end{document}